\newcommand\apjcls{1}
\newcommand\aastexcls{2}
\newcommand\othercls{3}

\newcommand\papercls{\aastexcls}
\documentclass[tighten, times, twocolumn]{aastex62}  

\if\papercls \apjcls
\usepackage{apjfonts}
\else\if\papercls \othercls
\usepackage{epsfig}
\usepackage{margin}
\usepackage{subcaption}
\usepackage{times}
\fi\fi
\usepackage{ifthen}
\usepackage{natbib}
\usepackage{amssymb, amsmath}
\usepackage{appendix}
\usepackage{etoolbox}
\usepackage[T1]{fontenc}
\usepackage{paralist}
\usepackage{tikz}
\usepackage{changepage}
\usepackage{soul}
\if\papercls \apjcls
\newcommand\aas{\ref@jnl{AAS Meeting Abstracts}}
\newcommand\dps{\ref@jnl{AAS/DPS Meeting Abstracts}}
\newcommand\maps{\ref@jnl{MAPS}}
\else\if\papercls \othercls
\usepackage{astjnlabbrev-jh}
\fi\fi

\if\papercls \aastexcls
\hypersetup{citecolor=blue, 
            linkcolor=blue, 
            menucolor=blue, 
            urlcolor=blue}  
\else
\usepackage[
bookmarks=true,           
bookmarksnumbered=true,   
colorlinks=true,          
citecolor=blue,           
linkcolor=blue,           
menucolor=blue,           
urlcolor=blue,            
linkbordercolor={0 0 1},  
pdfborder={0 0 1},
frenchlinks=true]{hyperref}
\fi

\if\papercls \othercls

\else

\fi

\providecommand{\adsurl}[1]{\href{#1}{ADS}}

\makeatletter
\patchcmd{\NAT@citex}
  {\@citea\NAT@hyper@{%
     \NAT@nmfmt{\NAT@nm}%
     \hyper@natlinkbreak{\NAT@aysep\NAT@spacechar}{\@citeb\@extra@b@citeb}%
     \NAT@date}}
  {\@citea\NAT@nmfmt{\NAT@nm}%
   \NAT@aysep\NAT@spacechar\NAT@hyper@{\NAT@date}}{}{}

\patchcmd{\NAT@citex}
  {\@citea\NAT@hyper@{%
     \NAT@nmfmt{\NAT@nm}%
     \hyper@natlinkbreak{\NAT@spacechar\NAT@@open\if*#1*\else#1\NAT@spacechar\fi}%
       {\@citeb\@extra@b@citeb}%
     \NAT@date}}
  {\@citea\NAT@nmfmt{\NAT@nm}%
   \NAT@spacechar\NAT@@open\if*#1*\else#1\NAT@spacechar\fi\NAT@hyper@{\NAT@date}}
  {}{}
\makeatother

\makeatletter
\DeclareRobustCommand{\lowcase}[1]{\@lowcase#1\@nil}
\def\@lowcase#1\@nil{\if\relax#1\relax\else\MakeLowercase{#1}\fi}
\makeatother

\DeclareSymbolFont{UPM}{U}{eur}{m}{n}
\DeclareMathSymbol{\umu}{0}{UPM}{"16}
\let\oldumu=\umu
\renewcommand\umu{\ifmmode\oldumu\else\math{\oldumu}\fi}

\if\papercls \othercls

\else

\fi

\let\oldsim=\sim
\renewcommand\sim{\ifmmode\oldsim\else\math{\oldsim}\fi}
\let\oldpm=\pm
\renewcommand\pm{\ifmmode\oldpm\else\math{\oldpm}\fi}
\newcommand\by{\ifmmode\times\else\math{\times}\fi}

\newbox{\wdbox}
\renewcommand\c{\setbox\wdbox=\hbox{,}\hspace{\wd\wdbox}}
\renewcommand\i{\setbox\wdbox=\hbox{i}\hspace{\wd\wdbox}}

\newcount\timect
\newcount\hourct
\newcount\minct
\newcommand\now{\timect=\time \divide\timect by 60
         \hourct=\timect \multiply\hourct by 60
         \minct=\time \advance\minct by -\hourct
         \number\timect:\ifnum \minct < 10 0\fi\number\minct}

\catcode`@=11

\newcommand\comment[1]{}

\newcommand\commenton{\catcode`\%=14}

\renewcommand\math[1]{$#1$}
\newcommand\mathshifton{\catcode`\$=3}

\let\atab=&
\newcommand\atabon{\catcode`\&=4}

\let\oldmsp=\sp
\let\oldmsb=\sb
\def\sp#1{\ifmmode
           \oldmsp{#1}%
         \else\strut\raise.85ex\hbox{\scriptsize #1}\fi}
\def\sb#1{\ifmmode
           \oldmsb{#1}%
         \else\strut\raise-.54ex\hbox{\scriptsize #1}\fi}
\newbox\@sp
\newbox\@sb
\def\sbp#1#2{\ifmmode%
           \oldmsb{#1}\oldmsp{#2}%
         \else
           \setbox\@sb=\hbox{\sb{#1}}%
           \setbox\@sp=\hbox{\sp{#2}}%
           \rlap{\copy\@sb}\copy\@sp
           \ifdim \wd\@sb >\wd\@sp
             \hskip -\wd\@sp \hskip \wd\@sb
           \fi
        \fi}
\def\msp#1{\ifmmode
           \oldmsp{#1}
         \else \math{\oldmsp{#1}}\fi}
\def\msb#1{\ifmmode
           \oldmsb{#1}
         \else \math{\oldmsb{#1}}\fi}

\def\supon{\catcode`\^=7}

\def\subon{\catcode`\_=8}

\def\supsubon{\supon \subon}

\newcommand\actcharon{\catcode`\~=13}

\newcommand\paramon{\catcode`\#=6}

\comment{And now to turn us totally on and off...}

\newcommand\reservedcharson{ \commenton  \mathshifton  \atabon  \supsubon 
                             \actcharon  \paramon}

\catcode`@=12
\reservedcharson

\if\papercls \apjcls

\else

\fi

\newcommand\tnm[1]{\tablenotemark{#1}}

\newcommand\chisq{\ifmmode{\chi\sp{2}}\else\math{\chi\sp{2}}\fi}
\newcommand\redchisq{\ifmmode{ \chi\sp{2}\sb{\rm red}}
                    \else\math{\chi\sp{2}\sb{\rm red}}\fi}
\newcommand\Teq{\ifmmode{T\sb{\rm eq}}\else$T$\sb{eq}\fi}
\newcommand\mjup{\ifmmode{M\sb{\rm Jup}}\else$M$\sb{Jup}\fi}
\newcommand\rjup{\ifmmode{R\sb{\rm Jup}}\else$R$\sb{Jup}\fi}
\newcommand\msun{\ifmmode{M\sb{\odot}}\else$M\sb{\odot}$\fi}
\newcommand\rsun{\ifmmode{R\sb{\odot}}\else$R\sb{\odot}$\fi}
\newcommand\mearth{\ifmmode{M\sb{\oplus}}\else$M\sb{\oplus}$\fi}
\newcommand\rearth{\ifmmode{R\sb{\oplus}}\else$R\sb{\oplus}$\fi}

\shorttitle{Form factor in Quasar Main Sequence}
\shortauthors{Panda {\em et al.}}
\def\mbh{$M_\mathrm{BH}$\/}
\def\feii{Fe{\sc ii}}
\def\lledd{$L/L_\mathrm{Edd}$\/}
\def\rfe{$R_\mathrm{FeII}$}
\def\kms{km s$^{-1}$}
\newlength\imagewidth
\newlength\imagescale
\begin{document}

\title{The Quasar Main Sequence explained by the combination of Eddington ratio, metallicity and orientation}

\author[0000-0002-5854-7426]{Swayamtrupta Panda}
\affiliation{Center for Theoretical Physics (PAS), Al. Lotnik{\'o}w 32/46, 02-668 Warsaw, Poland}
\affiliation{Nicolaus Copernicus Astronomical Center (PAS), ul. Bartycka 18, 00-716 Warsaw, Poland}
\author[0000-0002-6058-4912]{Paola Marziani}
\affiliation{INAF-Astronomical Observatory of Padova, Vicolo dell'Osservatorio, 5, 35122 Padova PD, Italy}

\author[0000-0001-5848-4333]{Bo\.zena Czerny}
\affiliation{Center for Theoretical Physics (PAS), Al. Lotnik{\'o}w 32/46, 02-668 Warsaw, Poland}


\correspondingauthor{Swayamtrupta Panda}
\email{spanda@camk.edu.pl}


\begin{abstract}
{We address the effect of orientation of the accretion disk plane and the geometry of the broad line region (BLR) as part of an effort to understand the  distribution of quasars in optical plane of the quasar main sequence. We utilize the photoionization code CLOUDY to model the BLR incorporating the grossly underestimated form factor ($f$). Treating the aspect of viewing angle appropriately, we confirm the dependence of the \rfe\ sequence on \lledd\ and on the related observational trends - as a function of the SED shape, cloud density and composition, verified from prior observations. Sources with \rfe\ in the range 1 -- 2 (about 10\% of all quasars, the so-called extreme Population A [xA] quasars) are explained as sources of high, and possibly extreme Eddington ratio along the \rfe\ sequence. This result has important implication for the exploitation of xA sources as distance indicators for Cosmology. \feii\ emitters with \rfe$ > 2$ are very rare (<1\% of all type 1 quasars). Our approach also explains the rarity of these highest \feii\ emitters as extreme xA sources, and constrains the viewing angle ranges with increasing H$\beta$ FWHM. 
}
 \end{abstract}


\section{Introduction}
\label{introduction}


Active galactic nuclei (AGN) are accreting black holes, where quasars represent the high luminosity tip of the AGN population. Until the early 1990s, the analysis of spectral data for AGN was clouded by sample size and by correlation analyses, often providing inconclusive or contradictory results \citep{sul00}. Sample selection further adds to this challenge. It was only with the landmark principal component analysis (PCA) by \cite{bg92} that a framework with reproducible, systematic trends between quasar spectral parameters was presented.  The eigenvector 1 of the original PCA of \cite{bg92} gave rise to the concept of the quasar main sequence (MS)   \citep{sul00,Mar01}. The MS is customarily represented by a trend in the plane's full width half maximum (FWHM) of the high-ionized Balmer line H$\beta$ vs parameter \rfe, defined by the ratio between the integrated \feii\ over the range 4434-4684 \AA\ and  the H$\beta$ intensity. In addition to \rfe\ and FWHM(H$\beta$), several multi-frequency parameters involving, for example, the CIV blueshift, the soft and hard photon indices, the low-ionization emission line profiles, are found to be correlated with the MS (for a more exhaustive list of MS correlates see \citealt{fb17}). 

{From a theoretical viewpoint, a quasar spectrum can be modelled using 4 basic ingredients:  black hole mass (\mbh), Eddington ratio (\lledd), spin and viewing angle ($\theta$). The accreting material has angular momentum leading to the formation of a planar, disk-like structure. Observed spectral properties therefore depend on the viewing angle at which each source is seen. The presence of a dusty, molecular torus limits the viewing angle to $\sim$ 60 deg., based on the AGN unification scheme \citep{antonucci93,urrypadovani95,marin14,netzer2015}.  


In our recent works (\citealt{panda_frontiers,panda18b,panda19}) we were successful in modelling almost the entire MS diagram constructed for $\sim$20,000 SDSS quasars \citep{s11,sh14} using mainly the contribution from two of the aforementioned physical parameters: the black hole mass and the Eddington ratio (at a fixed viewing angle and zero spin). The viewing angle was fixed since the range of the viewing angles was restricted by the presence of the torus. This was made to be consistent of picking only the unobscured sources, also known as Type-1 quasars where the observer has an unimpeded view of the central core. However, it has been mentioned quite often that the viewing angle persistently affects the dispersion of the main sequence and is coupled with the effects of the other parameters \citep{sh14}. In addition to the viewing angle, other factors affect the location of a quasar along the MS. The parameter \rfe\ is dependent on the metal content, increasing with increasing metallicity. At high FWHM-low \rfe\ MS end, UV spectra indicate solar or even sub-solar metallicity (see \citealt{punslyetal18} for a case study). Supersolar metallicity is expected in the general population of quasars \citep{hamannferland92,nagaoetal06,warneretal04,shinetal13,sulenticetal14}. At the high \rfe\  end of the MS, previous studies suggested high metallicity, even above ten times solar \citep[e.g.,][]{baldwinetal03,negreteetal12,martinez-aldamaetal18,panda18b,panda19}.

An additional trend is a systematic increase in hydrogen density toward the high \rfe\ sources \citep{willsetal99,aokiyoshida99}. Observationally, the trend is associated with a very high prominence of the CIII] 1909\AA~ emission line; and by its (almost) complete disappearance in strong \feii\ emitters \citep[e.g.,][and references therein]{negreteetal12}. The strongest \rfe\ emitters (\rfe$>$1, called extreme Population A [xA]) are therefore characterized by high metallicity, a dense, low-ionization BLR. The presence of a low-ionization, stable, dense region which remains virialized even at the highest values of $L$ and \lledd\  \citep{sulenticetal17,vietrietal18} led to the consideration of these sources as possible  ``Eddington standard candles" \citep{wangetal13,ms14}. } The aim of this paper is to  account for  \rfe\ values in each spectral type along the MS, in a way consistent with the observational trends in metallicity, density, and spectral energy distribution (SED). Section \ref{method} describes the method applied in predicting the \feii\ strength and Section \ref{results} infers constraints on the relative frequency of the MS sources as a function of their spectral type (ST). 
\par
The paper is organized as follows. In Section \ref{method}, we describe the modeling with the photoionisation code CLOUDY \citep{f17} taking into account the explicit dependence of the viewing angle. In our earlier works, we generalized the entire MS following the dependences on each parameter separately. Here, we also study the co-dependence of the various physical parameters (Eddington ratio, cloud density, metallicity and SED) while modeling each spectral type on a case-by-case basis. In Section \ref{results}, we explain the outcomes of the photoionisation modeling emphasizing the agreement to the observations based on prevalences of sources in each spectral type. In Section \ref{discussions} and \ref{sec:conclusions} we provide motivation for extension of this work that will account for intra-cloud dynamics and its connection with the cloud metallicity.

\section{Methods}
\label{method}

The range of values of \rfe\ in each spectral bin is defined according to Fig. \ref{fig:ms}, following \citet{sulenticetal02}. 

\subsection{Cloudy simulations}
\label{sec:cloudy}

\begin{figure}[h]
\centering
\includegraphics[width=.475\textwidth]{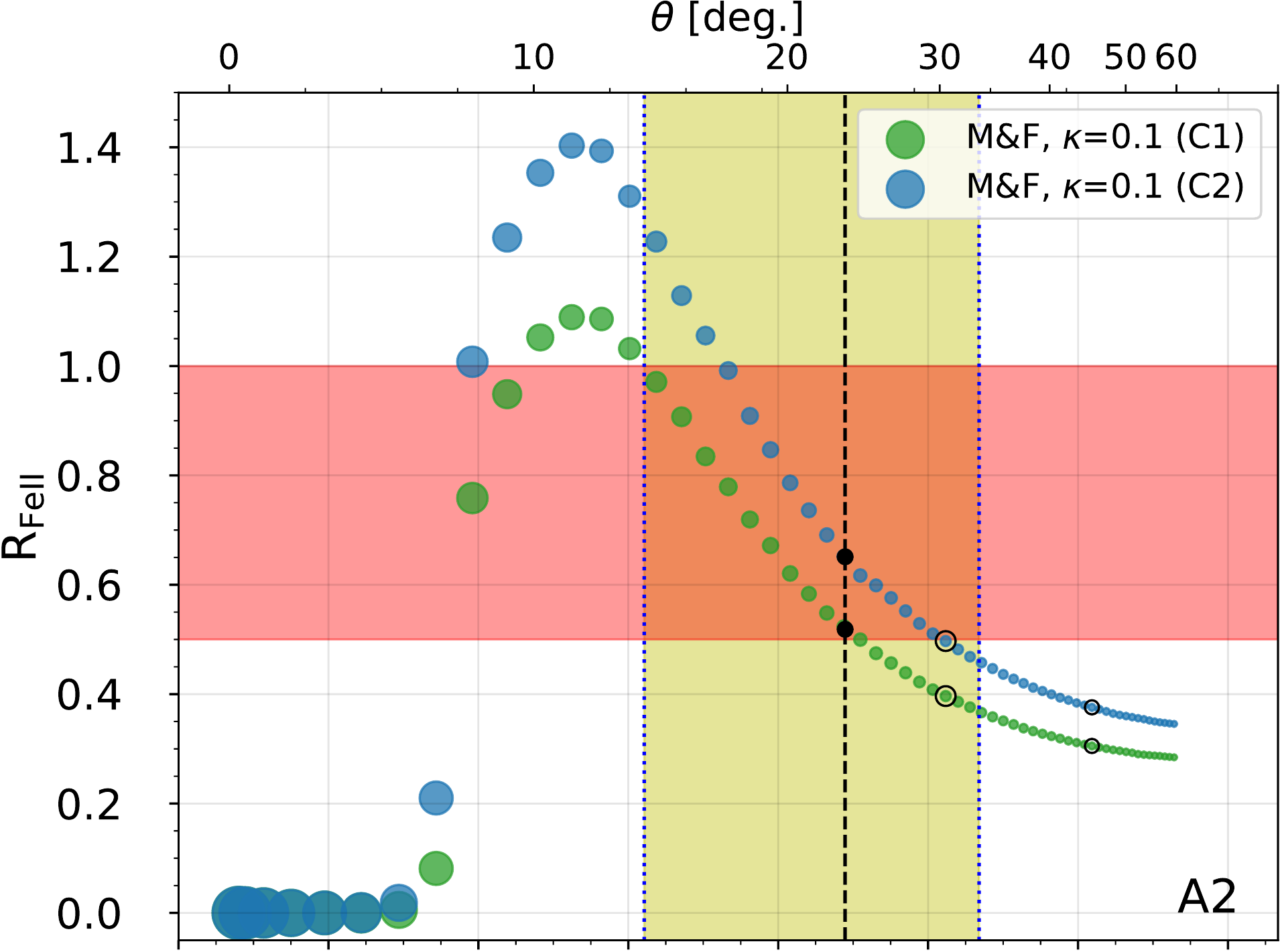}

\vspace{0.25cm}
\includegraphics[width=.475\textwidth]{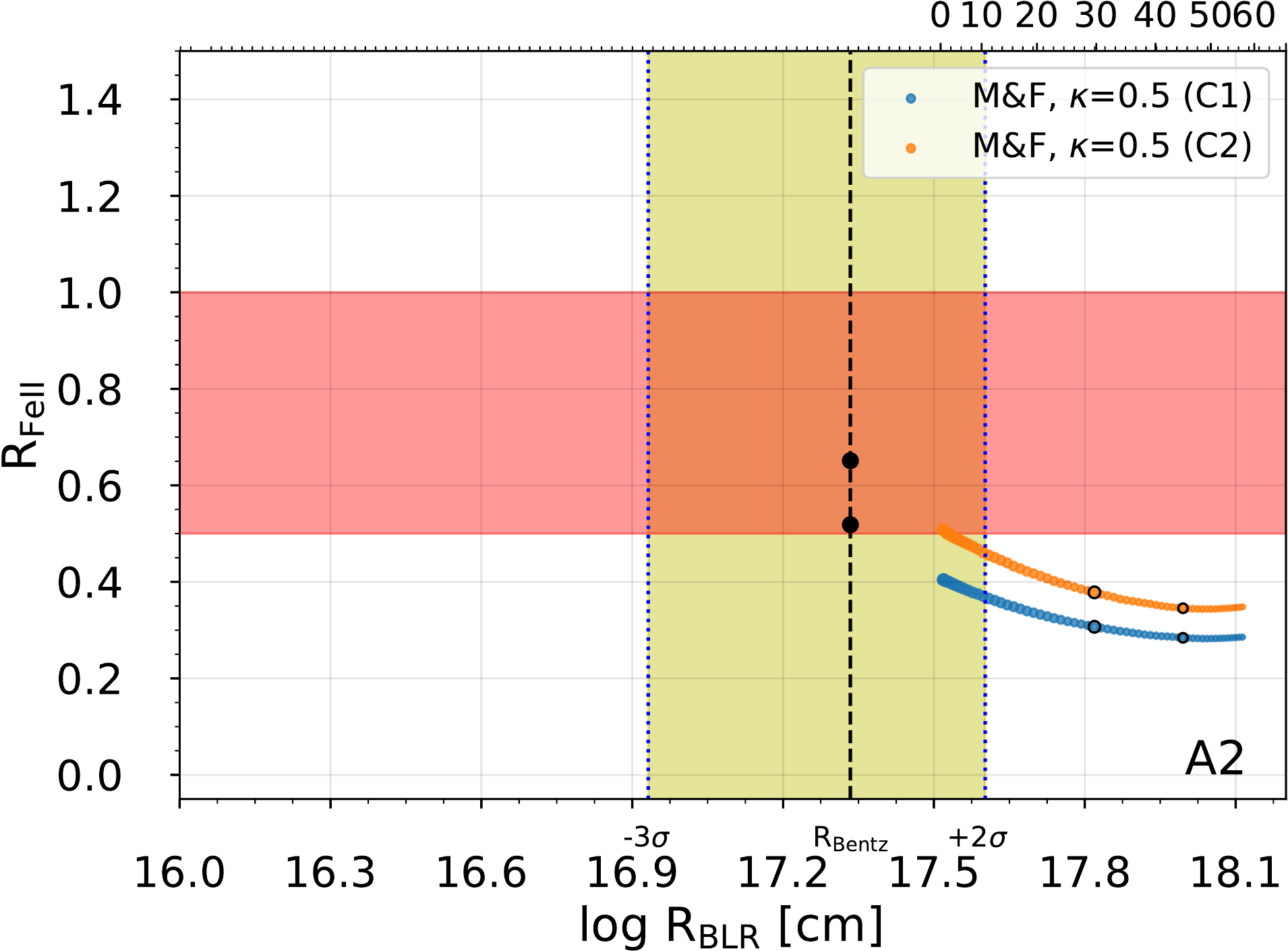}
\caption{A cut-out of the representation shown in Figure \ref{fig:ms}. The $\log R_{\mathrm{BLR}}$ vs. \rfe\ distribution as a function of increasing viewing angle [0, 60 deg.] from CLOUDY simulations is shown for two cases, C1 and C2, for two values of $\kappa$ i.e., 0.1 (\textbf{top panel}) and 0.5 (\textbf{bottom panel}). The size of the symbols is related to form factor (\textit{f}) that is dependent on the viewing angle ($\theta$) (see Equation \ref{eq:f}). Open circles mark the \rfe\ values expected for $\theta=30^{\mathrm{o}}$\ and $\theta=45^{\mathrm{o}}$. The horizontal red patch constrains the \rfe\ from the observation (see inset plot in Figure \ref{fig:ms}) for the spectral type A2. The dashed vertical line in black shows the radius for the BLR as predicted from the standard $R_{\mathrm{BLR}} - L_{5100}$ relation \citep{bentz13}. The black dots shows the agreement with the  $\log R_{BLR}$ - \rfe\ distribution for a theoretical 2-component SED (see \citealt{panda18b}) for the two cases. The vertical green patch shows the asymmetric range [$-3\sigma,+2\sigma$] accounting for the dispersion in the standard $R_{\mathrm{BLR}} - L_{5100}$ relation \citep{dupu2014,dupu2016,grier17,dupu2018}.}
\label{fig:A2_single}
\end{figure}

\begin{figure*}[tb]
    \centering
    \pgfmathsetlength{\imagewidth}{\linewidth}%
    \pgfmathsetlength{\imagescale}{\imagewidth/524}%
    \begin{tikzpicture}[x=\imagescale,y=-\imagescale]
        \node[anchor=north west] at (0,0) 
        {\includegraphics[height=0.74\imagewidth,width=\imagewidth]{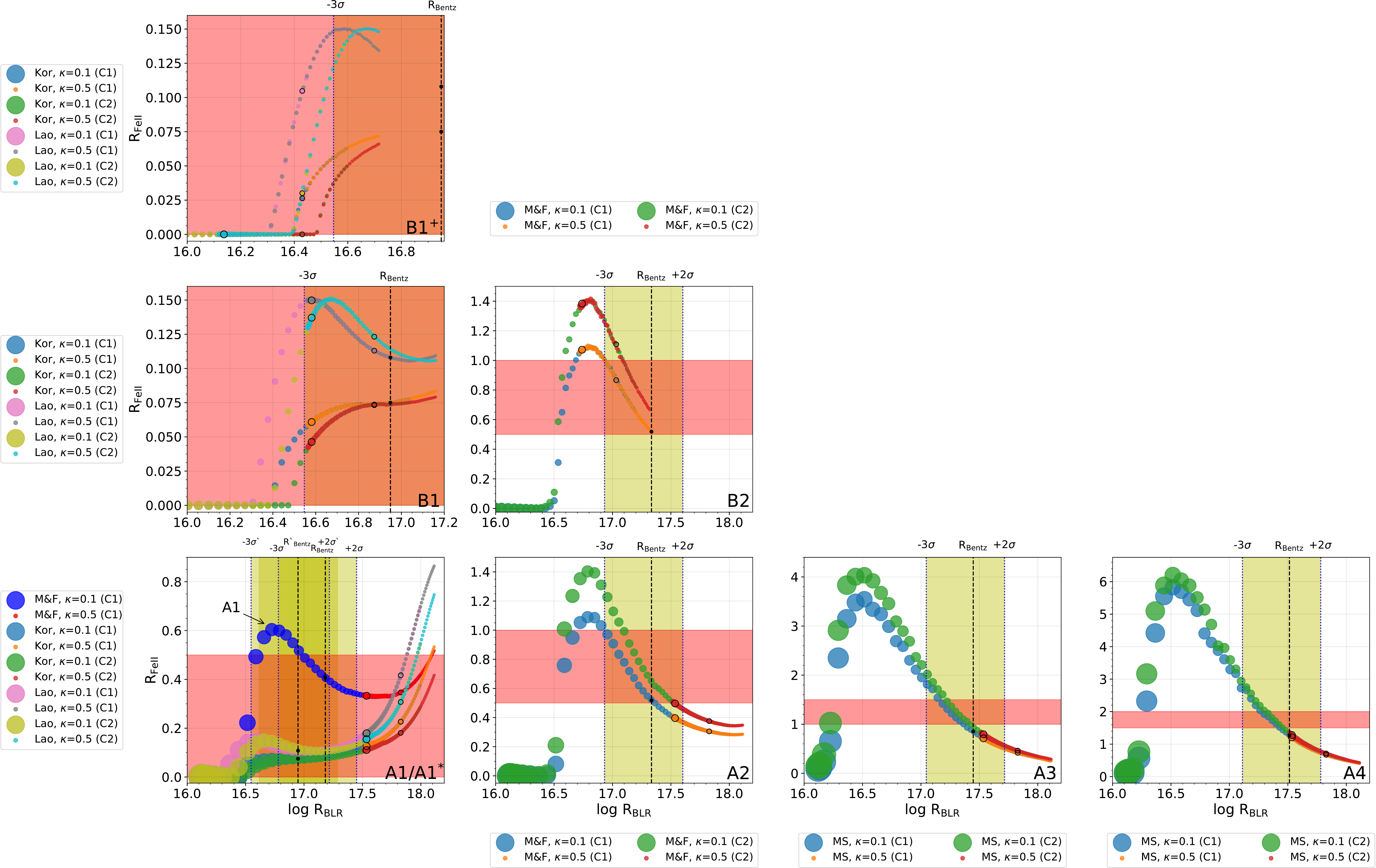}};
        \node[anchor=north west] at (300,5) {\frame{\includegraphics[height=0.425\imagewidth,width=0.425\imagewidth]{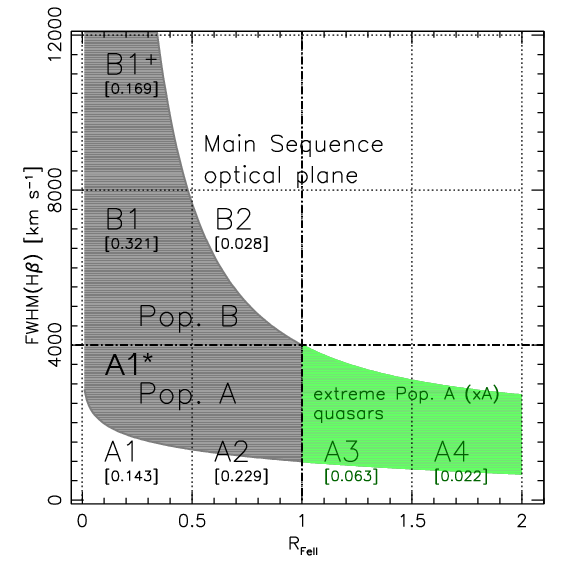}}};
    \end{tikzpicture}
    \caption{Results from a set of CLOUDY simulations performed for a constant density single BLR cloud assuming $\mathrm{M_{\mathrm{BH}}=10^8\;M_{\odot}}$, for cases C1 and C2 as reported in Tab. \ref{table1}. The plots are shown for the spectral bins (as defined in the inset panel) showing the distribution of changing \feii\ strength with changing BLR sizes computed from the virial relation. Average values of FWHM are used for each spectral bin. The computations are performed for viewing angle range 0--60 degrees (shown with increasing dot sizes), for a continuum SED from \citet[][{\tt M\&F}]{mf87}, \citet[][{\tt MS}; the one labeled NLSy1 in Fig. 12 of their paper]{ms14}, \citet[][{\tt Kor}]{kor97} and \citet[][{\tt Lao}]{laor97} for the respective spectral bins. The size of the symbols is related to form factor (\textit{f}) that is dependent on the viewing angle ($\theta$) (see Equation \ref{eq:f}). The black filled circle in each bins are the corresponding \feii\ strength at a radius constrained by \cite{bentz13} $R_\mathrm{BLR}$-$L_{\mathrm{5100}}$ relation shown here only for case C1. The parameters used for the different cases are given in Table \ref{table1}. Open circles mark the \rfe\ values expected for $\theta=30^{\mathrm{o}}$\ and $\theta=45^{\mathrm{o}}$. The color patches (in red) in each spectral bin denote the range of \rfe\ values as expected from observational evidences \citep{phillips78,bg92,grupe04,zamfiretal10,s11}; for extreme \rfe\ sources \citep{bergeron1980,lawrence1988,sulentic1990,lipari1991,lipari93,lipari94a,lipari94b}). The respective upper (+2$\sigma$) and lower (-3$\sigma$) bounds are shown by blue dashed lines about the $R_\mathrm{BLR}$ values estimated from the \cite{bentz13} relation (shown by black dashed lines) and the range is shown as green shaded regions. The inset diagram shows the optical plane of the Eigenvector 1, FWHM(H$\beta$) vs. \rfe. The shaded area indicatively traces the distribution of a quasar sample from \citealt{zamfiretal10}, defining the quasar main sequence. The thick horizontal dot-dashed line separates populations A and B. The vertical dot-dashed line marks the limit for extreme Population A (xA) sources with \rfe $\gtrsim$ 1. The prevalence of each spectral bin in the low-$z$ sample of \citet{marziani13} are shown within square brackets which shows the fraction of the sources in the respective STs to the total quasar sample considered. These prevalence allow for 97.5\% coverage in the considered STs.}
    \label{fig:ms}
\end{figure*}

The photoionization code CLOUDY \citep{f17} is used to relate \rfe\ parameter to the physical conditions in the BLR. We assume a constant density, single cloud model and the integrated optical \feii\ emission strength is parameterized by - \textit{black hole mass, Eddington ratio, cloud density and metallicity}, over a range of viewing angles from 0 to 60 degrees\footnote{The range of viewing angles  from 0 to 60 degrees is considered for the CLOUDY computations and this "full" range is shown in Fig \ref{fig:A2_single}, \ref{fig:ms} and \ref{fig:bh}. But, for the purpose of estimating the prevalences in each spectral types, the range of viewing angle is limited within 0 to 45 degrees (inclusive). This restriction is made to include unobscured sources in the quasar main sequence.}, where the viewing angle is defined as the angle between the axis perpendicular to the disc and the line of sight to the observer. The upper limit in the viewing angle is intended to exclude obscured sources in accordance with unification schemes \citep{antonucci93,urrypadovani95}. In the new version of CLOUDY, the FeII emission is modelled with 371 levels upto 11.6 eV, including 68,535 transitions based on the FeII model of \citet{verner99} \citep{f13,f17}, which shows quite good agreement with many observational FeII templates in the optical (\citealt{bg92,vc2003,kovacevic2010}). We consider two cases (C1 and C2) of parameterization systematically varying along the sequence (Table \ref{table1}). The sequence itself is divided into spectral sub-types as in \citet{marziani13}. For spectral types A, we assume fixed FWHM = 2000 \kms, for B1 = 6000 \kms, for B2 = 5000 \kms, and for B1$^+$ = 10000 \kms.  The spectral type  A1$^*$ is defined with the same physical conditions of  B1 and is  meant to represent intrinsic Pop. B sources seen at low $\theta$ \citep{marinantonucci16}, and assumes FWHM = 2000 \kms\, as for the A bins. The abundances are estimated using the \textit{GASS10} module \citep{gass10} in CLOUDY. The effect of micro-turbulence to model the MS has been shown to be of importance \citep{panda18b,panda19}, where the optical plane of quasars is indeed positively affected by inclusion of modest values of micro-turbulence (there is a $\sim$ 50\% increase in the \rfe\ when the turbulent velocity is increased to 10-20 \kms. Increasing the turbulence beyond 20 \kms\ reduces the net \rfe\ and for 100 \kms\ the value approaches close to the \rfe\ value obtained for the case with no turbulence). This micro-turbulence factor has not been accounted here for simplicity and will be addressed in a future project. 
\par
\begin{figure}[tb]
\centering
\includegraphics[width=\columnwidth]{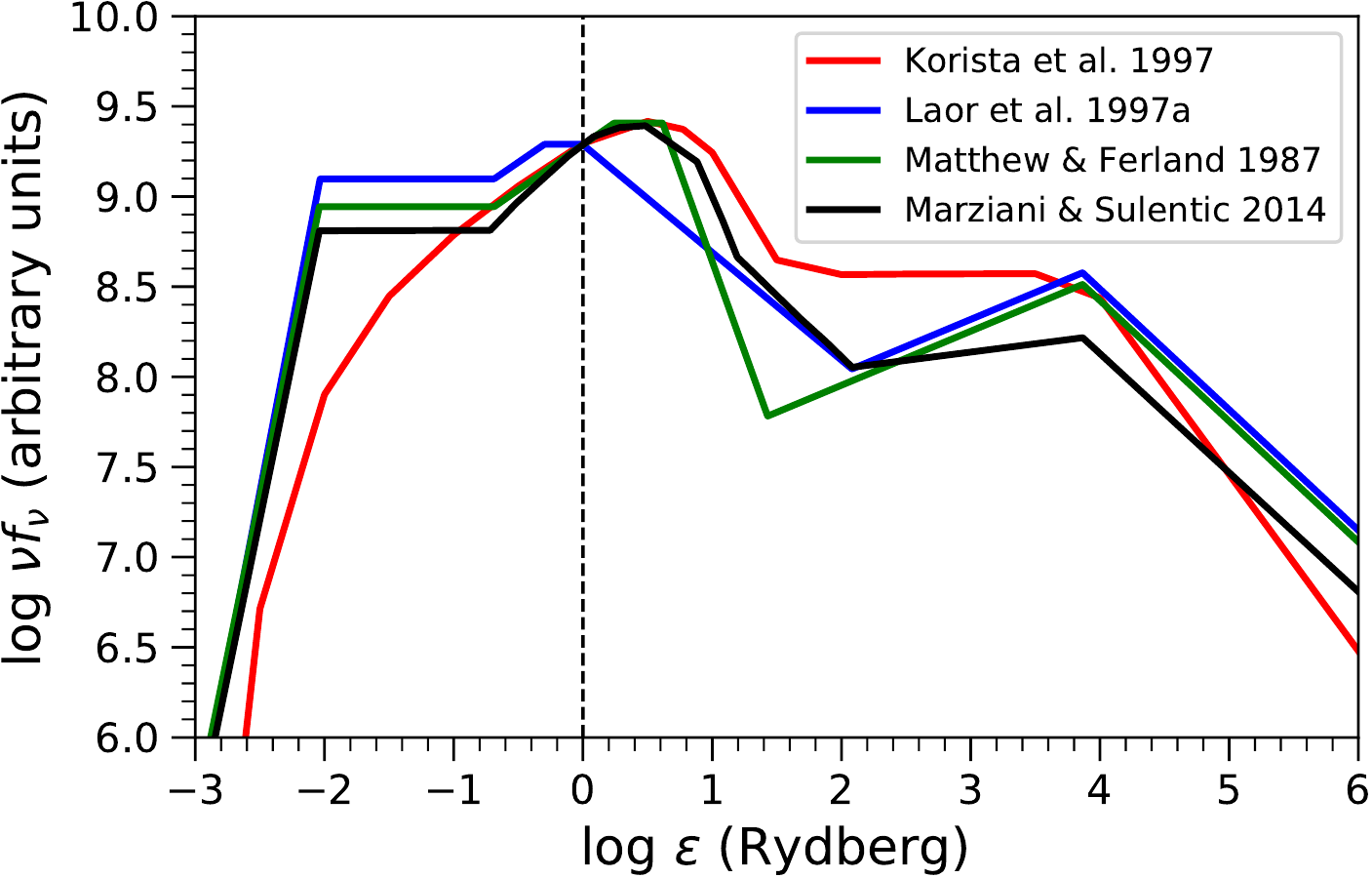}
\caption{Continuum spectral energy distributions (SEDs) used in this paper. The distributions are shown in red \citep[][]{kor97}, blue \citep[][]{laor97}, green \cite[][]{mf87} and  black \cite[][]{ms14}. The emitted power in arbitrary units is plotted as a function of the photon energy in Ryd. The SEDs have been normalised at $\log\epsilon$ = 0 which is shown with a black dashed line.}
\label{fig:sed}
\end{figure}

 
Figure \ref{fig:sed} shows the different SEDs which have been incorporated in this modelling. The SEDs have been normalised at $\log \epsilon = 0$ (where $\epsilon$ is the photon energy in Rydbergs) to highlight the differences in their shapes. For the low-FWHM, low \rfe\ spectral types e.g. A1-A2, we utilise the continuum shape defined by \citet{mf87} appropriate for Pop. A quasars. The SED from \citet{ms14} is utilised for low-FWHM, high \rfe\ spectral types e.g. A3-A4, which is appropriate for highly accreting quasars. For Population B sources, we have taken the continuum shape defined from \citet{kor97} and \citet{laor97}, as the exact behaviour of the continuum is still a work in progress. The assumption of the \citet{kor97} and \citet{laor97} SEDs brackets the relatively large spread in the SED, as well as the possibility that the disk may see a harder continuum than the observer. The \citet{laor97} SED is representative of the observed SED for sources that primarily occupy the Pop. B bins in the MS, possibly down to A1* ($\gtrsim$ 2000 \kms).  
 The  differences in SED shapes are particularly significant between 1 - 25 Rydberg    around the optical-UV bump where, on the low-energy side galaxy absorption prevents the observations and, on the high energy side the uncertainties about the Comptonization in the disk atmosphere and in the warm/hot corona are large. \citet{laor97} has a low prominence of the Big Blue Bump (blue line) in Figure \ref{fig:sed} where the Big Blue Bump (\citealt{czerny87, richards06}) refers to  the optical-UV peak in the SED which represents the emission from the accretion disk.  \citet{kor97} found that the BELRs perceive a harder continuum due to an increase in  soft-ionizing photons ($f_{\nu} \propto \nu^{-2}$, 13.6 - 100 eV). This can be seen in Figure \ref{fig:sed}, where the \citet{kor97} curve (red line) shows a significantly higher spectral energies, for $\log\;\epsilon \sim (1,4)$, compared to the other continua.


\begin{deluxetable*}{lcccccccccccccc}
\tabletypesize{\footnotesize}
\tablecolumns{15}
\setlength{\tabcolsep}{2pt}
\renewcommand{\arraystretch}{0.825}
\tablecaption{Photoionization models of spectral types and associated prevalences\tablenotemark{*}  \label{table1}}
\tablehead{\colhead{Case  }                          &
           \colhead{ST}                      &   \colhead{$Z$}    &
           \colhead{$\log n_\mathrm{H}$}                           &
           \colhead{$L/L_\mathrm{Edd}$}                     & 
          \colhead{SED}& 
         \colhead{$\theta$\tablenotemark{a}}&
          \colhead{$\log R_\mathrm{BLR}$\tablenotemark{b}}&
          \multicolumn{5}{c}{$\tilde{n}$} \\  \cline{9-13}
        &   & [$Z_{\odot}$]   & [cm$^{-3}$] & & & [degrees]  & [cm] & A1 &A2 &A3 &A4 &A5f\tablenotemark{c}  
          } 
\startdata
C1/C2  &  A1  & 5 & 10.5 & 0.2 & M\&F & 0 -- 45 & $16.12 - 17.83$ &   0.92 & 0.08 & 0.00 & 0.00 & 0.00   \\
C1/C2  &  A1  & 5 & 10.5 & 0.2 & M\&F & $10.9 - 26.8$  &  $16.78 - 17.45$ &   0.26 & 0.05 & 0.00 & 0.00 & 0.00 \\
C1  &  A2  & 5 & 11  & 0.5 &M\&F &0 -- 45 & $16.12 - 17.83$   & 0.735 & 0.215 & 0.05 & 0.00 & 0.00\\
C1  &  A2  &  5 & 11 & 0.5 &M\&F & $13.51 - 32.7$  & $16.93 - 17.60$     & 0.25 & 0.20 & 0.00 & 0.00 & 0.00 \\
C2  &  A2  & 7.5 & 11  & 0.5 &M\&F &$0 -  45$ &  $16.12 - 17.83$    & 0.58   &  0.31  & 0.11  &  0.00 & 0.00\\
C2  &  A2  &  7.5 & 11 & 0.5 &M\&F &  $13.5 - 32.7$   & $16.93	- 17.60$    & 0.09  & 0.30  & 0.05  & 0.00 & 0.00  \\
C1  &  A3  & 10 & 12 & 1  & MS &  0 -- 45  & $16.12 - 17.83$   & 0.255 & 0.455 & 0.125 & 0.06 & 0.105  \\
C1  &  A3  & 10 & 12 &  1  & MS & $15.9  - 38.5$  &  $17.05 - 17.72 $  &  0.00 & 0.45 & 0.12 & 0.04 & 0.00  \\
C2  &  A3  & 12.5 & 12 & 1  & MS &  0 -- 45  & $16.12 - 17.83$   &  0.12  & 0.54    & 0.14  & 0.07& 0.13   \\
C2  &  A3  & 12.5 & 12 & 1  & MS &  15.9 - 38.5  &  $17.05 - 17.72 $ & 0.00&   0.40 & 0.14 & 0.07 &  0.00 &  \\
C1  &  A4   & 20  & 12 &1.5  &MS & 0 -- 45 & $16.12 - 17.83$      & 0.00 & 0.425 & 0.23 & 0.115 & 0.23  \\
C1  &  A4  & 20 & 12 &1.5  &MS & $17.2 - 42.0$ & $17.11 - 17.78 $      & 0.00 & 0.30 & 0.23 & 0.12 & 0.08   \\
C2  &  A4   & 22.5  & 12 &1.5  &MS & 0 -- 45 & $16.12 - 17.83$      &   0.00   &  0.38  & 0.25  & 0.12 &  0.25  \\
C2  &  A4   & 22.5  & 12 &1.5  &MS & $15.9 - 38.5$ & $17.11 -	17.78$  &  0.00  & 0.25  & 0.25 & 0.12  & 0.11     \\
C1  & A1$^{\ast}$&  0.5 &10 & 0.05 &Lao/Kor &0 -- 45 & $16.12 - 17.83$    &  1.00 & 0.00 & 0.00 & 0.00  & 0.00\\ 
C1  & A1$^{\ast}$&  0.5 &10 & 0.05 &Lao/Kor &  $7.5 - 19.9$    & $16.55 - 17.22$  &  0.175 & 0.00 & 0.00 & 0.00  & 0.00\\ 
C2  & A1$^{\ast}$&  1.0 &10 & 0.075 &Lao/Kor &0 -- 45 & $16.12 - 17.83$ &   1.00 & 0.00 & 0.00 & 0.00  & 0.00\\ 
C2  & A1$^{\ast}$&  1.0 &10 & 0.075 &Lao/Kor &  $7.5 - 19.9$         & $16.55 - 17.22$ & 0.175 & 0.00 & 0.00 & 0.00 & 0.00 \\ 
 \hline
& & & & & & &  &   B1 & B2 & B3 & B4 & B5f\tablenotemark{d} \\
\hline 
C1  & B1 & 0.5 & 10 & 0.05  &Kor &16 -- 45 & $16.10 - 16.87$   & 0.87  & 0.00 & 0.00 & 0.00 & 0.00   \\
C1  & B1 & 0.5 & 10 & 0.05  &Kor    & $37.4 - 45$ &   $16.55 - 17.22$   & 0.30 & 0.00 & 0.00 & 0.00 & 0.00    \\
C1  & B1 & 0.5 &10  & 0.05  &Lao & $12 - 45$ & $16.55 - 17.22$   & 0.93  & 0.00 & 0.00 & 0.00 & 0.00   \\
C1  & B1 & 0.5 & 10 & 0.05  &Lao & $37.4 - 45$ & $16.55 - 17.22$      & 0.30 & 0.00 & 0.00 & 0.00 & 0.00   \\
C2  & B1 & 1.0 & 10 & 0.075  &Kor &$18 - 45$ & $16.10 - 16.87$  & 0.83  & 0.00 & 0.00 & 0.00 & 0.00   \\
C2  & B1 & 1.0 & 10 & 0.075  &Kor & $37.4 - 45$  & $16.55 - 17.22$  & 0.30 & 0.00 & 0.00 & 0.00 &0.00 \\
C2  & B1 & 1.0 & 10  & 0.075  &Lao & $13 - 45$ & $16.55 - 17.22$   & 0.91  & 0.00 & 0.00 & 0.00 & 0.00   \\
C2  & B1 & 1.0 & 10 & 0.075  &Lao &    $37.4 - 45$   &$16.55 - 17.22$ &  0.30 & 0.00 & 0.00 & 0.00 & 0.00 \\
C1  & B1$^{+}$&0.5 & 10 & 0.05 &Kor  & $28 - 45$  & $16.08 - 16.43$  & 0.60 & 0.00  & 0.00 & 0.00 & 0.00   \\ 
C1  & B1$^{+}$ & 0.5 & 10 & 0.05  &Kor & \ldots\tablenotemark{e}  &  $16.55 -  17.22$  & 0.00 & 0.00  & 0.00 & 0.00 & 0.00    \\
C1  & B1$^{+}$&0.5 & 10 & 0.05 &Lao  &  $21 - 45$ & $ 15.86 - 16.42 $  & 0.77 & 0.00  & 0.00 & 0.00 & 0.00   \\ 
C1  & B1$^{+}$&0.5 & 10 & 0.05 &Lao  & \ldots\tablenotemark{e} &    $16.55 - 17.22$   &  0.00 & 0.00  & 0.00 & 0.00 & 0.00      \\ 
C2  & B1$^{+}$&1.0 & 10 & 0.075 &Kor  & $31 - 45$  & $16.08 - 16.43$  & 0.51 & 0.00  & 0.00 & 0.00 & 0.00   \\ 
C2  & B1$^{+}$ & 1.0 & 10 & 0.075  &Kor &  \ldots\tablenotemark{e}  &  $16.55 - 17.22$   & 0.00 & 0.00  & 0.00 & 0.00 & 0.00  &  \\
C2  & B1$^{+}$&1.0 & 10 & 0.075 &Lao  &  $24 - 45$ & $ 15.86 - 16.42 $  & 0.70 & 0.00  & 0.00 & 0.00 & 0.00   \\ 
C2  & B1$^{+}$&1.0 & 10 & 0.075 &Lao    &    \ldots\tablenotemark{e}     & $16.55 - 17.22$   & 0.00 & 0.00  & 0.00 & 0.00 & 0.00    \\ 
C1  & B2   & 5 & 11 & 0.5 & M\&F & $8  - 45$ & $15.42 - 17.03$ &   0.25 & 0.35 & 0.37 & 0.00 & 0.00  \\ 
C1  & B2   & 5 & 11  & 0.5 & M\&F &  $38.3 - 45$   & $ 16.93 -	17.60$ &  0.00 & 0.23 & 0.01 & 0.00  & 0.00 \\ 
C2  & B2 & 7.5 & 11  & 0.5 & M\&F & $8 - 45$ & $15.42 - 17.03$ &  0.24 & 0.04 & 0.69 & 0.00 & 0.00  \\ 
C2  & B2 & 7.5  & 11 & 0.5 & M\&F &  $38.9  -        45$ & $ 16.93 -	17.60$ &  0.00 & 0.00  & 0.24 & 0.00 & 0.00 \\ 
\enddata
\tablenotetext{*}{Parameters and prevalences are shown only for the case with $\kappa$ = 0.1.}
\tablenotetext{a}{Viewing angle range. Here, the full range refers to 0 to 45 degrees.}
\tablenotetext{b}{$R_\mathrm{BLR}$ range associated with the $\theta$\ range following Eq. \ref{eq:mbh}.}
\tablenotetext{c}{\rfe $>$ 2, which would correspond to ST A5 and beyond. For A5f, \rfe\ $\geq$ 2.5.}
\tablenotetext{d}{The \rfe\ range for B5f is same as A5f.}
\tablenotetext{e}{No viewing angles allowed within the $R_\mathrm{BLR}$ range.}
\end{deluxetable*}

\subsection{Effect of viewing angle}
\label{sec:viewing_angle}

The virial equation 
\begin{equation}
M_\mathrm{BH} = \frac{ R_\mathrm{BLR} v_\mathrm{K}^2}{G}
\end{equation} 
allows us to estimate the mass of super-massive black holes from the rotational velocity $v_\mathrm{K}$ of the line emitting gas assuming it is in circular Keplerian orbits. The velocity $v_\mathrm{K}$ can be obtained from the extent of broadening of spectral line profiles due to the motion of the line emitting gas under the effect of the central potential. In practice, the $v_\mathrm{K}$ is obtained from its radial velocity projection - the line FWHM. The  line FWHM can be written as 
${\mathrm{FWHM}^{2}}/{4} =  v^{2}_\mathrm{iso} +  v^{2}_\mathrm{K}\sin^{2}\theta$,  where $v_\mathrm{iso}$ is the isotropic velocity component, and is therefore related to the ``true'' Keplerian velocity by $v_\mathrm{K}^{2} = f \mathrm{FWHM}^{2}$.   This basically translates into the effective determination of the form factor ({also known as the virial factor or structure factor})  
\begin{equation}
 f = {1}/{4[\kappa^2 + \sin^2\theta]}, 
 \label{eq:f}
\end{equation}
where $\kappa = v_\mathrm{iso}/v_\mathrm{K}$. In other words the emitting gas is confined in a flattened distribution viewed at an angle $\theta$. The validity of the expression for $f$\ is supported by several lines of evidence \citep[e.g.,][]{collin2006,mejia-restrepoetal18a,negreteetal18}.

The virial mass takes the form 
\begin{equation}M_\mathrm{BH} = f\frac{ R_\mathrm{BLR}  \mathrm{FWHM}^2}{G} = \frac{R_\mathrm{BLR} \mathrm{ FWHM}^{2}}{G{(4 \cdot (\kappa^{2} + \sin^{2} \theta))}}
\label{eq:mbh}
\end{equation} 

The Eddington ratio $L/L_\mathrm{Edd} = L/({\mathcal L_{0}}M_\mathrm{BH})$ can be  correspondingly written as 

\begin{equation}
\frac{L}{L_\mathrm{Edd}}  = \frac{L}{\mathcal L_{0}} 
\frac{G (4 \cdot (\kappa^{2} + \sin^{2} \theta)) }{R_\mathrm{BLR} \mathrm{ FWHM}^{2}} \label{eq:lledd}
\end{equation}
where, $\mathcal L_{0} = 1.249\times 10^{38}\;\mathrm{ergs\;s^{-1}}$. The inclusion of the orientation dependent form factor ($f$) makes it possible to restrict ``permitted" ranges in BLR radii.
\par 

\section{Results}
\label{results}

For the convenience of the readers, we show a representative example (Figure \ref{fig:A2_single}) of the result shown in Figure \ref{fig:ms}. Here, we focus on the spectral type A2, according to the classification by \citet{marziani13}. We assumes a black hole mass $10^8 M_{\odot}$. We use the SED from \citet{mf87} as appropriate for the chosen spectral class. Then, we fix the local density and the metallicity for the ionised gas cloud. We assume a specific Eddington ratio and compute the bolometric luminosity ($L$). The monochromatic luminosity (at 5100 \AA) is then estimated using the normalisation coefficient as a function of the black hole mass and the mass accretion rate (see Equation 5 in \citealt{panda18b}). The code requires to specify the inner radius and the column density that is used to estimate the size of the ionized cloud to self-consistently solve the radiative transfer through the medium. To estimate the distance of the cloud from the central ionizing source, we use the virial relation which is a function of the inclination angle (see Section \ref{sec:viewing_angle}). The cloud column density is fixed at $10^{24}$ cm$^{-2}$ \citep{panda18b,panda19} in all of our models. To estimate the abundances we utilize the \textit{GASS10} model \citep{gass10} which is incorporated in CLOUDY. We have assumed two cases for $\kappa$ (=$v_\mathrm{iso}/v_\mathrm{K}$): 0.1 for a keplerian-like distribution, and 0.5 for a thick-disk representation \citep{collin2006}. In particular, we found that assuming a $\kappa$=0.1, provides a better agreement to the \rfe\ values corresponding to the observations. We tested two sets of cases, C1 and C2 (see Table \ref{table1}). In ST A2,  we used different values for the metallicity, keeping the cloud density, the Eddington ratio and the SED shape same in the two cases. We found, in the ST A2, that increasing the metallicity from 5Z$_{\odot}$ to 7.5Z$_{\odot}$ enhances the FeII emission by $\sim 25\%$. This eventually affects the prevalences that are estimated. 
\par 
In a similar way, we now analyse all the spectra types of AGN along the Quasar Main Sequence. The specific parameters used for each spectra bin are summarized in Table \ref{table1} and Figure \ref{fig:ms}. We consider two sets of these parameters for the CLOUDY simulations attempting to reproduce the trends described in metallicity ($Z$), density ($n_{H}$), \lledd, and SED summarized in Section \ref{introduction}. C1 assumes a systematic increase in the first three parameters for the sequence of increasing \rfe\ (i.e., from A1 to A2, etc.). SED shapes assume a prominent big blue bump and a steeper X-ray spectrum in Pop. A. C2 assumes a similar trend as C1 but considers higher values of the physical parameters. Figure \ref{fig:ms} illustrates the diversity in the \feii\ strength as a function of the BLR radius for the respective spectral bins. Here, according to the selected parameter range, we obtain \rfe\ values that can be over 6. The highest values  depend on the selection of a suitable SED (e.g. \citealt{ms14}) combined with Eddington ratio values above 1,  for very dense BLR gas  ($10^{12}$\;cm$^{-3}$, see \citealt{panda18b,panda19}) with a composition that is few times solar. The inset of Fig. \ref{fig:ms} reports the prevalence of each spectral bin in the low-$z$ sample of \citet{marziani13} allowing for 97.5\% coverage in the considered STs.  Sources with \rfe$\gtrsim 2$ are rare in optically-selected samples such as the one of \citet{marziani13}, but  their prevalence depends on the selection criteria. They were found to be $\sim 1$\%\ of all quasars in SDSS based samples \citep{zamfiretal10,marziani13}, but to be  almost two third of the 76 soft-X bright sources of \citet{grupeetal99}. Of the 47 xA, 8 have \rfe $\gtrsim 2$, with a maximum value of almost 4.   While \rfe$\approx$6 requires maximization of density, Eddington ratio and Z, the physical condition for bin A3 in C2 (see Table \ref{table1}) are already sufficient to produce a significant fraction of sources with \rfe$\gtrsim 2$. 

{In each bin of Fig. \ref{fig:ms}, the curves representing each case as a function of $R_\mathrm{BLR}$\ predict  \rfe\ values that are changing because $R_\mathrm{BLR}$\ is affecting the ionization parameter. The \rfe\ values  are also a function of the viewing angle $\theta$, because of the coupling of the $\theta$ and $R_\mathrm{BLR}$\ in Eq. \ref{eq:mbh} (both mass and FWHM are fixed). In this framework, not all \rfe\ values are equiprobable, as the probability of observing an angle $\theta$\ is $p(\theta) \propto \sin \theta$. In each spectral bin of the Population A sequence, a fraction of the \rfe\ values is appropriate for different spectral types (for example, in bin A2 values of \rfe $< 0.5$ and \rfe $>1$ are possible, up to 1.4).  The last 5 columns of Table \ref{table1} report the ``distribution function" of the STs derived  for the physical conditions assumed for each original ST, computed by integrating $p(\theta)$ within the $\theta$\ limits, i.e., $ \tilde{n} = (\cos \theta_\mathrm{min} - \cos \theta_\mathrm{max})/n$, where $\theta_\mathrm{min}$\ and $\theta_\mathrm{max}$\ are set by the limits in $R_\mathrm{BLR}$\ as visible in Fig. \ref{fig:ms}, and $n \approx 0.293$.  The distribution among spectral types depends, for a given bin, on the case considered as well as on the limits on $R_\mathrm{BLR}$. The current model assumed a fixed FWHM per bin  and the resultant $R_\mathrm{{BLR}}$ range for the full range of viewing angles 0--45 degrees (Table \ref{table1}) is dependent on this. 

We can restrict $R_\mathrm{{BLR}}$ values to be consistent with the $R_\mathrm{{BLR}}$-$L_{5100}$\ relation \citep{bentz13} within a broad range meant to include the full range of $R_\mathrm{BLR}$\ at a given $L$  \citep{du2015, du2016, grier17, dupu2018} (+3$\sigma$ to -5.5$\sigma$, with $\sigma \approx 0.134$ dex). This condition basically covers almost all the permitted 0--45 degrees angle range. In Fig. \ref{fig:ms}, we consider a narrower range within {$+2\sigma$  and -3$\sigma$\ around $R_\mathrm{BLR}$} predicted by the \citet{bentz13} relation (shaded vertical strips in Fig. \ref{fig:ms}).  The asymmetric limits to the $R_\mathrm{{BLR}}$ account for the sources that have shown shorter time-lags and consequently deviate from the relation \citep{du2015, du2016, grier17, dupu2018}. To compute the $R_\mathrm{{BLR}}$ from the $R_\mathrm{{BLR}}$-$L_{5100}$\ relation, we used the \textit{Clean2} model from \citep{bentz13}, and utilised a normalization that is consistent with $\theta$ = 45 degrees to compute the optical monochromatic luminosity (at 5100\AA). The second lines of C1 and C2 listed for each ST in Table \ref{table1} report the $\theta$\ ranges associated with the restricted $R_\mathrm{{BLR}}$\ ranges, and the corresponding distribution function.

The physical conditions of A1 cannot be  assumed for all AGN, as they would predict 92\%\ of objects in bin A1 with a small fraction in A2 (8\%; first row of Table \ref{table1}), and the observed prevalence values reported in the bins of Fig. \ref{fig:ms} indicate that A2 is the most populated ST in the \rfe\ sequence along the Pop. A bins. No sources with \rfe $\gtrsim$ 1 would be possible, in contrast to observed prevalence.  On the other hand, the C1 and C2 physical parameters  account for the relatively high occupation in bin A1, and the smooth distribution between A1 and A2. However, if we assume the C1 condition for A3 in the $\theta$\ range 0--45 degrees, the prevalence A1/\-A2/A3\-/A4/A5f would as reported on the 7th row of Table \ref{table1}, by integrating over the probability associated with any viewing angle in the range 0--45 degrees.  {Keeping with C1, the integration over the probability within the strip allowed by  reverberation mapping over the probabilities of \rfe\ values for bin A2 and A3, assuming 50\%\ occurrence, produces a distribution function { A1/\-A2/A3\-/A4/A5f $\sim$ 0.28/0.59/0.10/0.03/0.00},  qualitatively consistent with the relative frequencies observed for Pop. A.}  }  The physical parameters of A3 in C2, apart from A4 in general, can already explain the very rare, strongest \rfe\ emitters.


In the B bins, the physical parameters assumed for C1 and C2  easily account for the typically weak \feii\ emission. The lower half of Table \ref{table1} shows that the parameters assumed for A1*, B1 and B1$^+$ do not predict any \rfe $>0.5$ emission. The FWHM(H$\beta$) assumed for B1 and B1$^+$ are 6000 and 10000 \kms\ respectively. The important consequence is -- the smallest values   of $\theta$\ corresponding to the symmetry axis almost aligned with the line of sight become impossible for the fixed \mbh\ = 10$^8$ M$_\odot$. This is not a serious problem for B1, as a sizable fraction of sources is still possible even if the restriction on $R_\mathrm{BLR}$\ is considered. On the contrary, for B1$^+$ no case is possible, and an \mbh\ increase is suggested. In other words, we can explain the vertical displacement in Fig. \ref{fig:ms} if sources are observed preferentially at higher $\theta$ for a given mass. At the same time, higher masses are needed to account for the broadest profiles. In actual samples, where the \mbh\ is not fixed, the vertical spread in the \rfe$<$0.5 section (upper-left quadrant in the inset plot of Fig. \ref{fig:ms}) of the MS is probably due to the combined effect of orientation and the spread in \mbh.  

\begin{figure}[tb]
\centering
\includegraphics[width=\columnwidth]{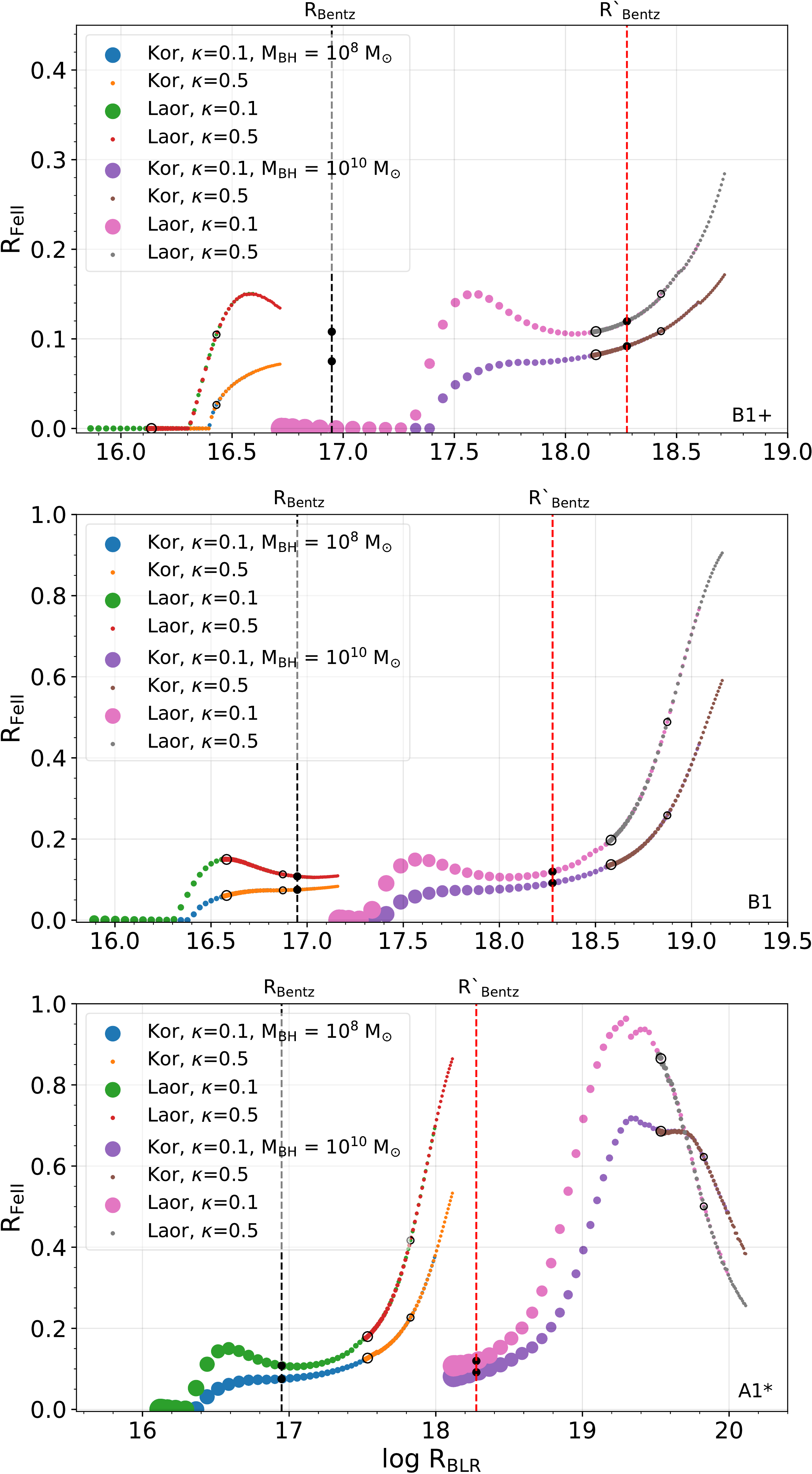}
\caption{Effect of black hole mass. Results from a set of CLOUDY simulations assuming $\mathrm{M_{\mathrm{BH}}=10^8\;M_{\odot}}$ and $\mathrm{M_{\mathrm{BH}}=10^{10}\;M_{\odot}}$. The plots are shown for the spectral bins A1, B1 and B1$^+$.  \rfe\ changes  along  with $R_\mathrm{BLR}$ computed from the virial relation (Eq. \ref{eq:mbh}), as in Fig. \ref{fig:ms}. Average values of FWHM are used for each spectral bin. The computations are performed for viewing angle range 0--60 degrees, for a continuum SED from \citep[][]{kor97} and \citep[][]{laor97} for the respective spectral bins. The size of the symbols is related to form factor (\textit{f}) that is dependent on the viewing angle ($\theta$) (see Equation \ref{eq:f}). The filled circles in black in each bins are the corresponding \feii\ strength at a radius constrained by \cite{bentz13} $R_\mathrm{BLR}$-$L_{\mathrm{5100}}$ relation for $\mathrm{M_{\mathrm{BH}}=10^8\;M_{\odot}}$ and $\mathrm{M_{\mathrm{BH}}=10^{10}\;M_{\odot}}$ shown as dashed lines in black and red, respectively. Only C1 parameters are shown here.}
\label{fig:bh}
\end{figure}

Figure \ref{fig:bh} shows the variation of the \rfe\ in the vertical spectral bins for the original case $M_{\mathrm{BH}} = 10^8\;M_{\odot}$, and $10^{10}\;M_{\odot}$. Here, we assume the SEDs consistent for normal Seyferts \citep{laor97,kor97} and the cloud properties as per C1 (see Table \ref{table1}). We find that increasing the $M_\mathrm{BH}$ to such masses (consistent for quasars in evolved systems)  increases the net \feii\ emission, enough to account for \rfe\ in B2 if the physical condition assumed are as for ST A1 (Tab. \ref{table1}; bottom panel of Fig. \ref{fig:bh}), but not as much as to create a significant population of B3 emitters.


{The B2 ST  includes a small fraction of objects ($\lesssim$3\%) in the \citet{marziani13} sample. Assuming a FWHM = 5000 \kms, B2 spectral properties may be explained  as due to higher  Eddington ratio  with respect to B1 (if there is a restriction on $R_\mathrm{BLR}$), with either  larger mass (which would produce, in addition to an increase in \rfe, a vertical displacement in FWHM $\propto \sqrt{M_\mathrm{BH}}$) or higher $\theta$.} The values of $\theta$\ consistent with FWHM = 5000 \kms\ and the $R_\mathrm{BLR}$ -- $L$\ are between 39 and 45 degrees (Table \ref{table1}). They are the ones with the highest probability of occurrence. Fig. \ref{fig:ms} shows that values  around 50-60 degrees (for~ C1) are also possible for the ST B1 since the posterior part of the curve extending till 60 degrees, still lies within the expected range of \rfe. This implies that the occupation of the B2 bin may be in part due to sources suffering significant extinction and reddening of the emitting line region and of the continuum (i.e. them appearing as type 1.5 and 1.8).

We also obtain the theoretical templates for the FeII pseudo-continuum using CLOUDY, which we have compared to observational templates for Mrk 335 and  I Zw 1. These sources belong to the spectral bins A1 and A3 respectively. We found that the theoretical templates show best agreement if a reasonable turbulence is applied\footnote{for Mrk 335: 0-10 km s$^{-1}$; I Zw 1: 40-50 km s$^{-1}$ with super-solar metallicities 1-2.5 Z$_{\odot}$ and 4.8-5.4 Z$_{\odot}$, respectively.}. The derived metallicities for I Zw 1  are consistent with the findings in  literature  suggesting highly super-solar values \citep[see e.g.,][]{negreteetal12} although they are lower if models with turbulence are considered. Thus, there exists some coupling between turbulence and metallicity estimates. We will address this issue in detail in a subsequent work.

\section{Discussion}
\label{discussions}

Using modelling by CLOUDY we successfully connected the properties of the sources in all spectral bins of the MS to the local conditions in their BLR. The allowed distance ranges of the BLR and the favoured viewing angles are consistent with expectations. Current study does not yet include the turbulent velocity and the full range of the black hole masses, which will allow to cover better extreme B spectral bins. 

The coupling between the local BLR properties like density and metallicity with the global ones, like Eddington ratio are well seen in the independent observational works discussed in Section 1, and in our modelling. The nature of this coupling is not yet well modelled, as it touches the issue of the time evolution of the nucleus and its surrounding. Perhaps vigorous starburst, leading to high metallicity, is indeed needed to power high Eddington ratio sources, although the possible time delay between the two episodes make the observational study of this issue rather complex.

\section{Conclusions}
\label{sec:conclusions}

A main result of the present investigation is the ability to explain the \rfe\ values along the main sequence, up to the highest observed values. The BLR radii are consistent with the one derived from reverberation mapping. We identify three possible physical condition: in spectral types B and A1, relatively low density, low $Z$ and low Eddington ratio, account for the weak \feii\ emission. At moderate \rfe, physical conditions appear consistent with the view of a moderate density ($n \sim 10^{11}$ cm$^{-3}$ \citep{matsuokaetal07,martinez-aldamaetal15}, intermediate Eddington ratio, and typical quasar metallicity. Sources with \rfe$\gtrsim$1 are accounted for by higher density, radiative output at Eddington limit, and high metallicity. These are the extreme properties that were inferred for some objects by previous work \citep{negreteetal12}. An extensive study will be performed which will incorporate a wider and more complete range of the parameter space from a multidimensional viewpoint. 

\section*{Acknowledgements}
The project was partially supported by the Polish Funding Agency National Science Centre, projects 2015/17/B/\-ST9/\-03436/ (OPUS 9), 2017/26/\-A/ST9/\-00756 (MAESTRO  9) and MNiSW grant DIR/WK/2018/12. PM acknowledges the Programa de Estancias de Investigaci\'on (PREI) No. DGAP/DFA/2192/2018 of UNAM, and funding from the INAF PRIN-SKA 2017 program 1.05.01.88.04. SP and PM would like to acknowledge the organisers of \href{https://matteobachetti.github.io/supereddington2018//}{``Breaking the limits 2018: Super-Eddington accretion onto compact objects''} where this project was first set up. We thank the anonymous referee for constructive suggestions to the manuscript that helped to reach the present form.
\software {CLOUDY v17.01 (\citealt{f17}); MATPLOTLIB (\citealt{hunter07})}
\bibliography{main}
\end{document}